\catcode`\@=11
\newif\if@fewtab\@fewtabtrue
{\count255=\time\divide\count255 by 60
\xdef\hourmin{\number\count255}
\multiply\count255 by-60\advance\count255 by\time
\xdef\hourmin{\hourmin:\ifnum\count255<10 0\fi\the\count255}}
\def\ps@draft{\let\@mkboth\@gobbletwo
    \def\@oddhead{}
    \def\@oddfoot
       {\hbox to 7 cm{$\scriptstyle Draft\ version:\ \draftdate$
       \hfil}\hskip -7cm\hfil\rm\thepage \hfil}
    \def\@evenhead{}\let\@evenfoot\@oddfoot}

\def\ceqno{\global\@fewtabfalse
    \ifcase\@eqcnt \def\@tempa{& & &}\or \def\@tempa{& &}
      \or \def\@tempa{&}
      \or\def\@tempa{}\fi\@tempa
{\rm(\theequation)}}

\def\aeqno#1{\global\@fewtabfalse
    \ifcase\@eqcnt \def\@tempa{& & &}\or \def\@tempa{& &}
      \or \def\@tempa{&}
      \or\def\@tempa{}\fi\@tempa
{\rm(\theequation,#1)}}

\def\label#1{\ifnum\draftcontrol=1
 \global\def\draftnote{$\scriptstyle #1$}\fi
 \@bsphack\if@filesw {\let\thepage\relax
   \def\protect{\noexpand\noexpand\noexpand}%
\xdef\@gtempa{\write\@auxout{\string
      \newlabel{#1}{{\@currentlabel}{\thepage}}}}}\@gtempa
   \if@nobreak \ifvmode\nobreak\fi\fi\fi
  \@esphack}

\def\alabel#1#2{\label{#1}\global\@fewtabfalse
    \ifcase\@eqcnt \def\@tempa{& & &}\or \def\@tempa{& &}
      \or \def\@tempa{&}
      \or\def\@tempa{}\fi\@tempa
{\hbox to 3cm{\phantom{\rm(\theequation,#2)}
\draftnote \hfil}\hskip -3cm {\rm(\theequation,#2)}}}

\def\clabel#1{\label{#1}\global\@fewtabfalse
    \ifcase\@eqcnt \def\@tempa{& & &}\or \def\@tempa{& &}
      \or \def\@tempa{&}
      \or\def\@tempa{}\fi\@tempa
{\hbox to 3cm{\phantom{\rm(\theequation)}
\draftnote \hfil}\hskip -3cm{\rm(\theequation)}}}

\def\eqnarray{\def\draftnote{{}}\global\@fewtabtrue
\stepcounter{equation}\let\@currentlabel=\theequation
\global\@eqnswtrue
\global\@eqcnt\z@\tabskip\@centering\let\\=\@eqncr
$$\halign to \displaywidth\bgroup\@eqnsel\hskip\@centering\@eqcnt\z@
  $\displaystyle\tabskip\z@{##}$&\global\@eqcnt\@ne
  \hskip 1\arraycolsep \hfil${##}$\hfil
  &\global\@eqcnt\tw@ \hskip 1\arraycolsep
$\displaystyle\tabskip\z@{##}$
\hfil  \tabskip\@centering&\global\@eqcnt\thr@@\llap{##}\tabskip\z@
\cr}

\def\endeqnarray{\@@eqncr\egroup
      \global\advance\c@equation\m@ne$$\global\@ignoretrue}

\def\@eqnnum{\hbox to 3cm{\phantom{\rm(\theequation)} \draftnote
                         \hfil}\hskip -3cm {\rm(\theequation)}}

\def\@@eqncr{\let\@tempa\relax
    \ifcase\@eqcnt \def\@tempa{& & &}\or \def\@tempa{& &}
      \or \def\@tempa{&}
      \or\def\@tempa{}
\fi\@tempa
\if@eqnsw
\if@fewtab\@eqnnum\fi
\stepcounter{equation}\fi\global
\@eqnswtrue\global\@eqcnt\z@\global\@fewtabtrue\cr}

\def\draftcite#1{\ifnum\draftcontrol=1#1\else{}\fi}

\def\@lbibitem[#1]#2{\item{}\hskip -3cm \hbox to 2cm
{\hfil$\scriptstyle\draftcite{#2}$}\hskip
1cm[\@biblabel{#1}]\if@filesw
     {\def\protect##1{\string ##1\space}\immediate
      \write\@auxout{\string\bibcite{#2}{#1}}}\fi\ignorespaces}

\def\@bibitem#1{\item\hskip -3cm \hbox to 2cm
{\hfil $\scriptstyle\draftcite{#1}$}\hskip 1cm
\if@filesw \immediate\write\@auxout
       {\string\bibcite{#1}{\the\value{\@listctr}}}\fi\ignorespaces}

\catcode `\@=11
\catcode `\@=12
\def\ga{\gamma}         \def\Ga{\Gamma}
\def\be{\beta}
\def\al{\alpha}
\def\ep{\epsilon}
\def\la{\lambda}        \def\La{\Lambda}
\def\de{\delta}         \def\De{\Delta}
\def\om{\omega}         
        \def\Sig{\Sigma}

              \def\CC{{\cal C}}

              \def\CL{{\cal L}}
       \def\CN{{\cal N}}       \def\CO{{\cal O}}
\def\CP{{\cal P}}

\def\qq{ \begin{eqnarray} }
\def\qqq{ \end{eqnarray} }
\def\non{ \nonumber }
\newcommand{\no}{\noindent}
\newcommand{\vs}{\vspace}

\newcommand{\p}{\partial}

\newcommand{\rmp}{\mathrm{p}}

\newcommand{\rmx}{\mathrm{x}}

\newcommand{\hf}{{_1\over^2}}

\def\draftdate{\number\month/\number\day/\number\year\ \ \ \hourmin }

\global\def\draftcontrol{0}
\catcode`\@=12

\documentclass[12pt]{article}

\usepackage{mathbbol}

\begin{document}

\begin{center}

{\Large{\bf{Fourier's law from Closure Equations}}}

\vs{0.5cm}

{\large{Jean Bricmont}}

UCL, FYMA, chemin du Cyclotron 2,\\ 
B-1348  Louvain-la-Neuve, Belgium\\

\vs{ 0.2cm}

{\large{Antti Kupiainen}}\footnote{Partially supported by the 
Academy of Finland
}

Department of Mathematics,
Helsinki University,\\
P.O. Box 4, 00014 Helsinki, Finland\\

\end{center}
\vs{ 0.2cm}

\vskip 0.3cm
\begin{center}
\end{center}
\vskip 0.2cm

\begin{abstract}

We give a rigorous derivation  of Fourier's law from a
system of closure equations for a nonequilibrium 
stationary state of a
Hamiltonian system of coupled 
oscillators subjected to heat baths on the boundary.
The local heat flux is proportional to the  temperature
gradient with a temperature dependent heat conductivity
and 
the stationary temperature exhibits
a nonlinear profile.
\end{abstract}

\date{ }

\vskip 0.3cm
\begin{center}
\end{center}

One of the simplest and most fundamental nonequilibrium
phenomena is heat conduction in solids. It is described by
a macroscopic equation, Fourier's law, which
states that a local temperature gradient
is associated with a flux of heat $\cal J$ which is proportional to the gradient:
\qq
{\cal J}(x) =- k(T(x)) \nabla T(x)
\label{(I1)}
\qqq
where the {\it heat conductivity} $k(T(x))$ is a function  only of
the temperature at $x$.

Despite its fundamental nature, a derivation
of Fourier's law from  first principles, or even within a suitable approximation, such as a Boltzmann type equation, lies well beyond what can be mathematically proven
  (for reviews on the status of this problem, see \cite{BLR}, \cite{LLP}
and \cite{Spohn}).  The quantities $T$ and $\cal J$ in (\ref{(I1)}) are macroscopic variables,
statistical averages of the variables describing the microscopic dynamics of
matter. A first principle derivation of (\ref{(I1)}) entails a definition of
$T$ and $\cal J$ in terms of the microscopic variables and a proof of the law in
some appropriate limit.

In this letter we outline a rigorous proof  \cite{BK} of Fourier's law
starting from a closure approximation of the equations for
 the nonequilibrium stationary state of a
Hamiltonian system subjected to boundary heat baths. The
physical  situation we have in mind is a slab of crystal of linear extension $N$  heated at the two ends by temperatures   $T_{1}$ and $T_{2}$.  In this case one would expect $\nabla T$ and
$\cal J$ to be $ \CO (1/N) $ and  (\ref{(I1)}) to hold
up to corrections of order ${o} (1/N)$. This is indeed what
we prove in our model together with a detailed description of the temperature
distribution in the bulk.

 Since (\ref{(I1)}) is a
macroscopic law it is expected to hold for a classical system as
well as for a quantum one and the quantum corrections are expected to be small except at low
temperatures. A classical toy model describing
the above situation which has been intensively discussed
in recent years is given by 
coupled oscillators organized on  a strip of
width $N$ in $d$-dimensional  cubic lattice $\mathbb{Z}^d$. The oscillators are indexed by lattice
points $x=(x_1,\dots ,x_d)$ with $0\leq x_1\leq N$ and
carry momenta and coordinates $(p_x,q_x)$. 
The dynamics of the oscillators consists of two parts: Hamiltonian
dynamics in the bulk and noise on the boundary modelling
heat baths at temperatures   $T_{1}$ and $T_{2}$. 

We consider a Hamiltonian of the form
\qq
H (q,p) = {1\over 2} \sum_{x\in V} p^2_{x} + {1\over 2} (q,\omega^2q) +
{\la \over 4} \sum_{x\in V} q^4_{x}
\label{2(2)}
\qqq
which describes a system of coupled anharmonic
oscillators with coupling matrix $\omega^2$, i.e. 
$
(q,\omega^2 q) = \sum q_{x} q_{y} \omega^2 (x-y)
$. The noise is specified by the Gaussian random variables
$\xi_x(t)$ at sites $x$ on the boundary with covariance
\qq
< \xi_{x}(t) \xi_{y}(t')> = 4\ga \delta_{xy} (T_{1} \delta_{x_10} + T_{2}
\delta_{x_1N})\delta(t-t') \equiv 2C_{xy} \delta(t-t').
\label{2(4)}
\qqq
 The (stochastic) dynamics is given by 
$\dot q_x=p_x$ and
\qq
\dot p_{x} = (-{\p H \over \p q_{x}} - \ga_{x} p_{x}) + \xi_{x}
\label{2(1)}
\qqq
where the friction is $
\ga_{x} = \ga (\delta_{x_10} + \delta_{x_1N})
$ (more precisely, (\ref{2(1)}) is a Ito stochastic differential equation).
These equations define a Markov process
$(q(t),p(t))$ and we are interested in the stationary states
for this process.

In the {\it equilibrium} case of equal  temperatures, $T_{1}=T_{2}=T$, an explicit stationary state
is given by the Gibbs state 
${Z}^{-1} e^{-\beta H (q,p)} dq dp$
of  the Hamiltonian $H$ with
inverse temperature $\beta = 1/T$.
When  $T_{1} \neq
T_{2}$ there is no such simple formula  and,
indeed, in our setup, even the existence of a stationary state is an open
problem. In the $d=1$ case of a finite chain of $N$ oscillators the
existence is proved  under conditions (see \cite{RB}  for a review of such results, first obtained in \cite{EPR1}) as well as the
convergence of the  Markov process  to this state as
$t\to \infty$.

Supposing that we have a stationary state, let us formulate the statement
(\ref{(I1)}).  Writing $H$ as a sum of local terms, each one pertaining to
a single oscillator:
$
H =\sum_{x\in \Lambda} H_{x},
$
one has up to noise terms
$
\dot H_{x} =\nabla \cdot j(x),
$
where the {microscopic heat current} $j(x)$ will depend on $p_{y}$ and $q_{y}$
for $y$ near $x$ provided the coupling matrix $\omega^2$ is
short ranged. 
Let also $t(x) =
{1\over 2} p_x^2$ be the kinetic energy of the oscillator indexed by $x$. Then, the
macroscopic temperature and heat current in eq. (\ref{(I1)}) are defined by
$
T(x) =< t(x)> _{\mu}$
 and ${\cal J}(x) =< j(x)  >_{\mu}$
where $ <\cdot>_{\mu} $ denotes expectation in the stationary
state.

The only rigorous results in our model as far as deriving 
(\ref{(I1)}) are for the harmonic case of quadratic $H$ \cite{RLL, SL}. 
In that case, Fourier's law {\it does not hold}: the current $j(x)$ 
is  $\CO(1)$ as
$N\to\infty$ whereas $\nabla T = 0$ except near the boundary.
If $\la\neq 0$
the law seems to hold
in simulations in all dimensions \cite{AK}. In an analogous 
momentum conserving model
conductivity seems anomalous in low dimensions: $k$ in (\ref{(I1)}) depends on $N$
as $N^\al$ in $d=1$ and logarithmically in $d=2$. It is a major challenge to
explain the $\al$ which, numerically, seems to be in the interval $[1/3,2/5]$
(see  \cite{LLP1}, \cite{NR}, \cite{russian} for theories on $\al$).

In this paper we suppose the stationary state exists and
we study its properties via its
correlation functions. Let us denote $(q_x,p_x) = (u_{1x},  u_{2x}) $, $\La (u^{\otimes 3})_{\al x} = -\la\de_{\al,2}q_x^3$, $(\Ga u)_{x} = (0,\ga_{x}p_{x})^T$ and $\eta = (0,\xi)^T$.
Then (\ref{2(1)}) becomes
\qq
\dot u(t) = \Bigl((A-\Ga)u + \La (u^{\otimes 3})\Bigr)  + \eta (t)
\label{3(1)}
\qqq
where
$
A = \left(\begin{array}{ccccc}
0 & 1\\
-\omega^2 & 0
\end{array}\right)
$.
Applying (\ref{3(1)}) to the stationary state correlation functions
\qq
G_{n} (x_{1}, ..., x_{n}) = < u_{x_{1}} \otimes ... \otimes u_{x_{n}}
> 
\non
\qqq
these are seen to be solutions of the Hopf equations
\qq
(A_{n} - \Ga_{n}) G_{n} + \La_{n} G_{n+2} + \CC_{n} G_{n-2} = 0.
\label{3(4)}
\qqq
where
$
A_{n} = \sum A_{x_i} $
and $ \Ga_{n}$ is defined similarly. $\La_{n}$ and  $\CC_{n}$ are
 linear operators involving the
quartic vertex and noise covariance
$C_{xy}$ of  (\ref{2(4)}) respectively. The equations (\ref{3(4)}) have the drawback that they do not ``close": to solve for $G_{n}$,
we need to know $G_{n+2}$. 

We will now introduce an approximation that will lead to a closed set of nonlinear equations for $G_2$. For this we note that
for $\la$ small, the equilibrium $T_{1} = T_{2}$ measure is close to Gaussian.
When $T_{1} \neq T_{2}$ we expect this to remain true and we look for a Gaussian approximation to equation (\ref{3(4)}) for small $\la$
by means of a {\it closure}, i.e. expressing the $G_{n}$ in terms of $G_{2}$. Let $G^c_{4}$ be the connected correlation function
describing deviation from Gaussianity. 
Then the first equation in the hierarchy (\ref{3(4)}) reads:
\qq
(A_{2} - \Ga_{2} + \Sig_{2}) G_{2} + \La_{2} G^c_{4} + \CC = 0
\label{3(7)}
\qqq
where
$
\Sig_{2} (G_{2}) G_{2} = \La_{2} \sum G_{2} \otimes G_{2}
$. The simplest closure would be to drop the $\La_{2} G_{4}^c$ from (\ref{3(7)}). This leads to a nonlinear equation for $G_{2}$. It turns out that the solution to
this equation is qualitatively similar to the $\la=0$ case, i.e. $G_{2}$ does
not exhibit a temperature profile nor a finite conductivity. The only effect
of the nonlinearity is a renormalization of $\omega$.

The next equation in the hierarchy becomes after some some algebra
\qq
(A_{4} - \Ga_{4} + \Sig_{4}) G^c_{4} + b (G_{2}) + \La_{4} G^c_{6} = 0,
\label{3(8)}
\qqq
where $G^c_{6}$ is the connected six point function,
\qq
\Sig_{4} (G_{2}) G^c_{4} &=& \sum(\La_{4}( G_{2} \otimes G^c_{4})
-G_{2} \otimes \La_{2}G^c_{4}).
\non
\qqq
and
\qq
b(G_{2}) = \sum_{p} \La'_{4} (G_{2} \otimes G_{2} \otimes G_{2}),
\non
\qqq
where $\La'_{4}$ has  $\La$ 
acting on all the three factors $G_{2}$. 

(\ref{3(7)}) and (\ref{3(8)}) yield an exact equation for the two point
function with the connected six point function as an
input. Our closure
approximation consists of dropping the $G^c_{6}$  term in 
(\ref{3(8)}) thereby yielding a
closed set of equations for $G_{2}$. For simplicity we also drop the operators 
$\Sig_{2}$, $\Sig_{4}$ and $\Ga_4$: these could be included in
our analysis, but
do not change the main structure that is due to the term $b(G_{2}) $.
Hence the closure equation we study is for $G=G_2$:
\qq
(A_{2} - \Ga_{2}) G + \CN (G) +  \CC = 0
\label{3(10)}
\qqq
with
\qq
\CN = - \La_{2} A^{-1}_{4} b (G).
\label{4(1)}
\qqq

To write this more concretely, let us introduce
 the matrices $Q_{xy}=<q_xq_y>$,  $P_{xy}=<p_xp_y>$  and  $J_{xy}=<q_xp_y>$. Clearly $\dot Q=J+J^T$ so the
(1,1)-component of (\ref{3(10)}) says $J_{xy}=-J_{yx}$ and
we can write
$G = \left(\begin{array}{cccc}
Q & J\\
-J & P
\end{array}\right)$.

Next,
anticipating translation
invariance in the directions orthogonal to the 1-direction
we  write
\qq
G(x,y) = \int e^{i{p}(x_1+y_1)+ik(x-y)}  G (p,k) dp dk
\label{5(5)}
\qqq
where  the integrals over $p$ and $k_1$  are 
 Riemann sums on a $\pi\over 2N$ lattice. The inverse
 of 
 $A_{4}$ is written as
\qq
-A^{-1}_{4} = \int^\infty_{0} e^{t A_{4}} dt = \int^\infty_{0} R(t)^{\otimes
4} dt
\non
\qqq
where $R(t)=e^{tA}$. In Fourier space the latter is
\qq
\widehat R (t,q)= {1\over 2} \sum_{s=\pm 1} e^{(is\om (q)-\ep)t}
\left(\begin{array}{cccc}
1 & -is\om (q)^{-1} \\
is\om (q) & 1
\end{array}\right),
\label{5(0)}
\qqq
Then some algebra yields the following expression for the nonlinear term
\qq
&& N (p, k) = \sum_{\bf s} \int d \nu (\sum_1^4 s_{i} \om (p_{i} + k_{i}) + i \ep )^{-1}
\prod^2_{i=1}  W_{ s_{i}} (p_{i},k_{i})
\cdot \left(\begin{array}{ccccc}
0 & 0 \\
1 & is_{4}\om (p_{4}+k_{4})
\end{array}\right)
\non
\\
&&s_{3} \om
(p_{3}+k_{3}) 
 \Bigl[\om
(p_{3}+k_{3})^{-2} \de (2p_{3})  W_{s_{4}} (p_{4}, k_{4})
- \om (p_{4}+k_{4})^{-2} \de (2p_{4}) W_{s_{3}}
(p_{3},k_{3})\Bigr]
\label{59a}
\qqq
where
\qq
d\nu &=& \de
(2p - \sum (p_{i}+k_{i})) \de (\sum (p_{i}-k_{i}))\de (p-k-p_{4}-k_{4}) d { \bf  p} d {\bf k},
\label{5(10)}
\qqq
and ${\bf k}=(k_i)_{i=1}^4$ and similarily for ${\bf p}$ and
${\bf s}$.
$W$ is the following combination 
\qq
W_{s} (p, q) = \widehat Q (p, q)
+ is\om (p+q)^{-1} \widehat J (p, q).
\label{Br7}
\qqq

Eq. (\ref{3(10)}) is a nonlinear set of equations
for the pair correlation functions of our model.  In \cite{BK} 
we have proven that they have a unique solution which we
now proceed to explain.
The 1,2-component of (\ref{3(10)})  (coming from ${_d\over^{dt}}<qp>$)  gives 
\qq
P=  \om(p, k)^2Q + \hf((J\Ga -\Ga J) - \CN_{12}
(p,k) - \CN_{12} (p,-k))
\label{5(13)}
\qqq
where $
\om(p, k)^2 = \hf (\om (p+k)^{2} + \om (p-k)^{2})
$.
Since the nonlinear term $\CN$ depends only on $Q$ and $J$
this expresses 
$P$  in terms of them. The rest of (\ref{3(10)}) then yields
 two equations for the two
unknown functions  $Q$ and $J$ which become
\qq
\de \om^2 
 (Q,
 J)^T
+ \CN(Q,J)+
(\Ga J+J\Ga,
\Ga P+ P\Ga)^T
=
(0,
 C)^T
\label{8(1)}
\qqq
where $
N (Q,J)=
(
\CN_{12} (p,-k) - \CN_{12} (p,k),
-\CN_{22} (p,k))^T$
and
$
\de\om^2(p, k) =  \om (p+k)^{2} - \om (p-k)^{2}
$.
An important property of $\CN$ is that, for all $T$,
$\CN (TQ_{0},0) = 0$ where 
$Q_0=\om^{-2}
$. For $\ga=0$ these form  a 1-parameter family of solutions of 
(\ref{8(1)}).
In the equilibrium case $T_{1}= T_{2}$ and  $\ga\neq 0$ only one of these persists,
namely the one with $T=T_{1}$. This is the analogue
in the closure equation
of  the true equilibrium Gibbs state which has
$Q=Q_0+\CO(\la)$. 

Since we are looking for a solution that is locally in $x_1$
close to this equilibrium we need to understand the linearization
of the nonlinear term $\CN$ at $G_0$. It turns out
that it is given by an operator which is
a multiplication operator in the slow variable $ p$:
\qq
\CN(TQ_0+\de Q,\de J)(p,k)=\CL_{p} (\de J(p,\cdot),
 \de Q(p,\cdot))^T(k).
\label{8(2)}
\qqq
$\CL_{p} $ is a matrix of operators $\CL_{ij}(p) $. Each
of these acts on functions of $k$ as a sum of a
multiplication and an integral operator
\qq
(\CL_{ij}(p)f)(k)=A(p,k)f(k)+\int B(p,k,k')f(k')dk'.
\label{8(2a)}
\qqq
The integral kernel $B(p,k,k')$ is of the form
\qq
 B(p,k,k')= &=&   \sum_{\bf s}\int \De
\Bigl(\sum^2_{i=1} s_{i}\om (k_{i}) +  s_{3}\om (k'+p) + s_{4} \om (p-k)\Bigr) 
\non\\
&&\cdot\de (k-k_{1}-k_{2}-k') \rho_{{\scriptsize \textbf{s}}} (k_{1},k_{2},k',k,p)
dk_1 dk_{2}
\label{8(21)}
\qqq
where $\De (x) = \de (x)$ or $\CP \left({1\over x}\right)$
and $\rho_{{\scriptsize
 \textbf{s}}}$ is a smooth function. $A(p,k)$ is given by a similar expression integrated over $k'$. 
 
 The integrand in  (\ref{8(21)})  represents {\it phonon scattering}.
 The delta functions impose momentum and energy conservation
 when $p=0$ (it turns out that only terms with $\sum s_i=0$ contribute) which
 is the point where our functions are peaked: 
the translation invariant equilibrium has support at
$p=0$ and the nonequilibrium solution will also have most of its mass 
in the neighborhood of this point. Thus it is important
to understand $\CL_{0} $. For parity reasons
$
 \CL_{ij} (0)=0$ for  $ i\neq j$ whereas $ \CL_{11} (0) $ is invertible. Invertibility of $\CL_0$
would then follow from invertibility of $ \CL_{22} (0) $. This, however,
is not the case:  $ \CL_{22} (0) $ 
has {\it two zero modes}. 

 One of them is easy to understand.  Since
$\CN (TQ_{0},0) = 0$ for all  $T$, taking derivative
with respect to $T$, one finds
$ \CL_{22} (0) \om^{-2}=0$.
There is, however, also a second zero mode:
$ \CL_{22} (0) \om^{-3}=0$.

While the first zero mode has to persist for the full Hopf equations
due to the one parameter family of Gibbs states that solve them
for $\gamma=0$, the second one is an artifact of the closure
approximation.  The phonon scattering described by the nonlinear term
conserves phonon energy, leading to the first zero mode, and also
phonon number, leading to the second one. The connected
six-point correlation function which was neglected in
the closure approximation would produce terms that
violate phonon number conservation and remove the second zero mode.
However, for weak anharmonicity its eigenvalue
would be close to zero and should be treated as some
perturbation of the present analysis.

The second zero mode leads one to expect that our equations have in the
$\gamma=0$ limit a two parameter family of stationary solutions
which indeed is the case. These are given by
\qq
{Q}_{T,A}(x,y)=T\int e^{ik(x-y)}(\om(k)^2 - A \om (k))^{-1} dk .
\label{536a}
\qqq
The second zero-mode  is proportional to 
the derivative of ${Q}_{T,A}$ with respect to $A$, at $A=0$.
We are then led to look for solutions in the form
\qq
Q(x,y)=Q_{T(\rmx),A(\rmx)}(x-y)+q(x,y),
\label{Q1}
\qqq
where
the first term is of {\it local equilibrium} form with slowly
varying temperature and ``chemical potential" profiles $T(\rmx)$
and $A(\rmx)$ and where  $q$ is a perturbation orthogonal to
the zero modes in a suitable inner product.

Projecting
equation (\ref{8(1)}) on the complementary subspace of the
zero modes yields a nonlinear equation for $J$ and $q$ with
an invertible linear part. It can be solved by fixed point methods in
a suitable Banach space and yields $J$ and $q$ as
functionals of $T$ and $A$. The heat and phonon number currents
are given in momentum space by
\qq
{\cal J}^\al (\mathrm{p}) = -i \int dk e^{-i\mathrm{p}/2}  
\om (\rmp/2,k)^\al \sin k_1J (\rmp/2,k)
\label{5(33)}
\qqq
for $\al=1,0$ respectively and are thus nonlinear functionals
of $\nabla T$ and $\nabla A$.
This relation is the precise form
of the Fourier law. In particular we get relation  (\ref{(I1)}) for the heat current,
 up to corrections ${\CO}({\la^2\Delta T\over N})$ (here $\Delta T=
 T_2-T_1$),
with the thermal conductivity given by:
$$\kappa(T(x))={c\over \la^2T(x)^2}.
$$
Projecting then
equation (\ref{8(1)})
to the two left zero modes of
$ \CL_{22} (0) $ yields two conservation laws
for the  currents which
become nonlinear (and nonlocal) elliptic equations for the functions
 $T(\rmx)$
and $A(\rmx)$ whose solutions give 
for the {\it inverse temperature} $\beta(x)= T(x)^{-1} $
a linear
profile:
$\beta(x)  = \beta_1 + {|x|\over N} (\beta_2 - \beta_1 )$,
with corrections of order $\CO(\Delta T \la^2 {|x|\over N} )$  or of higher order  in $N^{-1}$. 

The main technical assumptions we need to prove these claims
are smallness of $_1\over^N$,  $\la$ and $ \Delta T $.
For convenience we take also the coupling $\ga$ to the
reservoirs small,  as
$\ga = N^{-1+\al}$ for some small $\al>0$. The coupling $\ga$ is important
to fix the boundary conditions for the elliptic equations determining
$T$ and $A$, but, for this, one only needs it to be bigger than
$N^{-1}$. We also need for our analysis that $\omega^2$ is
sufficiently {\it pinning}; in fact, we will choose
$
\omega^2 = (-\Delta + m^2)^2,
$
with $m$ large enough,
i.e. in momentum space $\omega (k) = 2 \sum^d_{i=1} (1-\cos k_{i}) + m^2$.
This choice simplifies several estimates.

The most crucial assumption is that the space dimension has
to be at least three. This is because of the low regularity of
the collision kernels in eq.   (\ref{8(21)}).  To prove the
spectral properties of the linear operators (\ref{8(2a)})
we need compactness of the integral operator $B$
in a nice enough space. In three dimensions this
holds in a space of H\"older continuous functions
whereas in two dimensions this is not the case.
Even in three dimensions, the resulting solutions
have low regularity in momentum space which
translates into long range correlations in 
physical space.

We finish by comparing our equations to the standard
kinetic theory (see \cite{Spohn} for  a discussion of
the kinetic theory of phonon systems). In our setup,
the kinetic limit is a {\it scaling limit}. One rescales
$x_1$ to the unit interval and takes
 $N\to \infty$, $\la \to 0$, while keeping $R=N\la^2$ constant. 
Our equations then formally become, for $\ga=0$,
the following equation for the function $V(x,k)=\om Q(x,k)+ i J(x,k)$, where $x\in [0, 1]$:
\qq
\nabla \om(k) \nabla_x V(x,k)=RN(V)
\label{k1}
\qqq
with 
\qq
&&N(V)=
{9\pi^2\over 2} \int dk_1 dk_3dk_3  (\om(k) \om(k_1)\om(k_2)\om(k_3))^{-1} .\non\\
&&
\de ( \om(k)+ \om(k_1)-\om(k_2)-\om(k_3)) \de(k+k_1-k_2 -k_3) . \non\\
&&[V(k_1) V(k_2)V(k_3)-V(k)(V(k_1)V(k_2)+V(k_1)V(k_3)-V(k_3)V(k_3))]
\label{k2}
\qqq
where the integration is over $k_i\in [0, 2\pi]^d$.  

Equation $N(V)=0$ has a two parameter family of solutions 
corresponding to  (\ref{536a}):
$V_{T,A}(x,k)= \frac{T}{\om(k) -A}$.
Then, 
imposing, a posteriori,  boundary conditions on equation (\ref{k1}), of the form $T(0)=T_1$,
$T(1)=T_2$,  $A(0)=A_1$,  $A(1)=A_2$, one expects, for $R$ large, the solution of (\ref{k1})
to be approximately of the form $V_{T(x), A(x)}$, with    $T(x), A(x)$ 
solved from a scaling limit of our equation for them  that becomes
\qq
\partial_x(D(T(x), A(x)) (\partial_xT(x), \partial_xA(x))^T)= 0,
\label{k3}
\qqq
where  $D$  an explicit $2\times 2$ matrix obtained by taking the scalar product of $L^{-1}$ with suitable vectors related to $V_{T(x), A(x)}$ and orthogonal to the zero modes (see \cite{ALS}). Solving (\ref{k3}) with the prescribed boundary conditions then  gives the approximate solution $V_{T(x), A(x)}$ of (\ref{k1}).

From perturbation theory one expects the  full Hopf equations to
reduce in the kinetic limit to the equation (\ref{k2}).
Thus our closure equations should have
the same kinetic scaling limit as the full theory.
It should be emphasized that they are
 more general than the  kinetic equations. Indeed, we do not take any limit: $\la$, $\tau$ and $N$ are fixed and we prove that our solution has the expected properties, with precise bounds on the remainder, for $\la$ and $\tau$ small and $N$ large. From a mathematical point of view, the Boltzman equation  (\ref{k2}) has its own problems
 due to the continuum nature of the variable $x$. Since
 we work with $N$ finite, some of these unphysical problems are absent
 in our analysis.
 
 In conclusion, the closure equations provide an approximation
 to the full Hopf equations of the nonequilibrium state that
 allow us to rigorously show how the Fourier law emerges
 as the system size gets large. Moreover, this approximation
 should become exact in the
kinetic scaling limit. It should provide a starting point
for an eventual first principles proof of the Fourier law. 
 
\vspace*{3mm}

\no {\bf {Acknowledgments.}}
We thank Joel Lebowitz, Rapha\"el Lefevere, Jani Lukkarinen, Alain Schenkel  and Herbert Spohn
for useful discussions.  A.K.  thanks the Academy of Finland
for funding.

\end{document}